\documentclass[11pt]{article}

\usepackage[final]{acl}

\usepackage{times}
\usepackage{latexsym}
\usepackage{caption}

\usepackage[T1]{fontenc}

\usepackage[utf8]{inputenc}

\usepackage{microtype}

\usepackage{inconsolata}

\usepackage{graphicx}

\usepackage{amsmath}
\usepackage{algorithm}
\usepackage{algpseudocode}
\usepackage{amsmath}
\usepackage{amssymb}
\usepackage{multirow}
\usepackage{booktabs}
\usepackage{adjustbox}
\usepackage{bbm}
\usepackage{tabularx,array} 
\newcolumntype{Y}{>{\centering\arraybackslash}X} 

\title{Marking Code Without Breaking It: Code Watermarking for Detecting LLM-Generated Code}

\author{
\quad Jungin Kim$^{\star}$
\quad Shinwoo Park$^{\star}$
\quad Yo-Sub Han$^{\dagger}$ \\
Yonsei University, Seoul, Republic of Korea \\
\texttt{\small jungin.kim@yonsei.ac.kr, pshkhh@yonsei.ac.kr, emmous@yonsei.ac.kr} 
}

\newcommand{\correspondingfootnote}{
    \let\oldthefootnote=\thefootnote
    \renewcommand{\thefootnote}{}
    \footnotetext{$\star$ Authors equally contributed.}
    \footnotetext{$\dagger$ Corresponding author.}
    \let\thefootnote=\oldthefootnote
}

\begin{document}

\maketitle

\correspondingfootnote 

\begin{abstract}

Identifying LLM-generated code through watermarking poses a challenge in preserving functional correctness.
Previous methods rely on the assumption that watermarking high-entropy tokens effectively maintains output quality.
Our analysis reveals a fundamental limitation of this assumption: syntax-critical tokens such as keywords often exhibit the highest entropy, making existing approaches vulnerable to logic corruption.
We present \texttt{STONE}, a syntax-aware watermarking method that embeds watermarks only in non-syntactic tokens and preserves code integrity.
For rigorous evaluation, we also introduce \texttt{STEM}, 
a comprehensive metric that balances three critical dimensions: 
correctness, detectability, and imperceptibility.
Across Python, C++, and Java, \texttt{STONE} preserves correctness, sustains strong detectability, and achieves balanced performance with minimal computational overhead.
Our implementation is available at \url{https://github.com/inistory/STONE-watermarking}.

\end{abstract}

\section{Introduction}
\label{sec:introduction}

\begin{figure}[t]
  \centering
  \includegraphics[width=\linewidth]{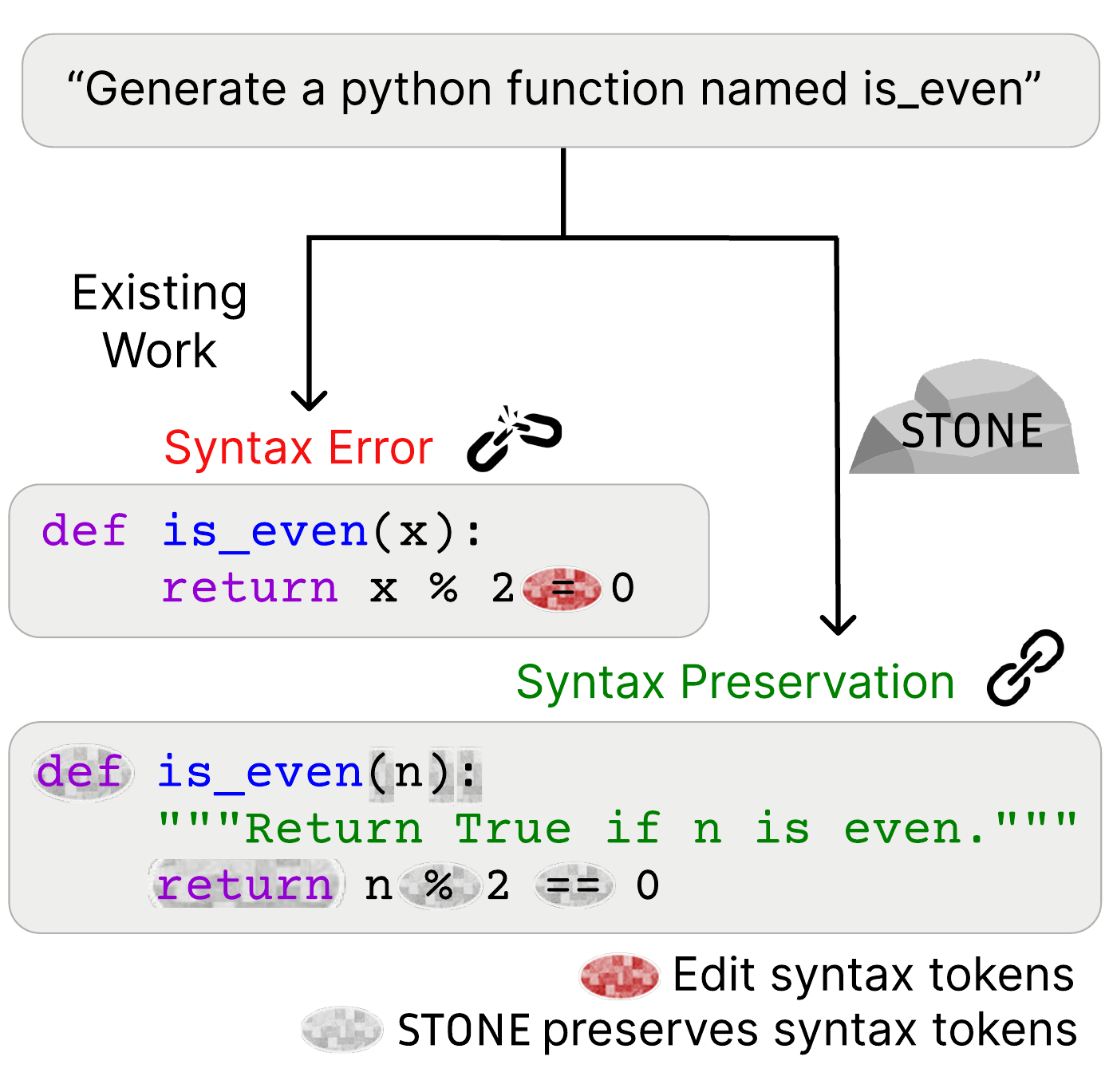}
\caption{
    Motivating example of \texttt{STONE} watermarking.
    Existing methods modify syntax tokens and cause syntax errors, 
    whereas \texttt{STONE} preserves syntax and embeds watermarks 
    without breaking code structure.
    }

  \label{fig:motivation}
\end{figure}

The rapid development of LLMs 
has significantly improved their ability to generate human-like code
~\citep{zan2022large,DBLP:conf/kbse/TangGLZXHL23,
DBLP:conf/sigsoft/0003X023}.
These advancements have unlocked new possibilities, 
including software development 
and automated code generation~\citep{nam2024using,wang2024teaching,Guo0XLTFWGB24}. 
However, this progress has raised 
new challenges in tracing 
the provenance of generated code~\citep{asurveyondetection,li2023protecting}. 
As LLM-generated code becomes 
increasingly indistinguishable 
from human-written code,
researchers have explored watermarking techniques 
to address this issue.
As illustrated in Figure~\ref{fig:motivation},
existing approaches often modify syntax-critical tokens, which can introduce syntax errors or alter program behavior.
In contrast, \texttt{STONE} preserves these tokens
and embeds watermarks only into non-syntactic elements,
thereby maintaining structural and functional integrity.
LLM watermarking embeds specific patterns 
into the generated output during the generation process 
to help identify its 
origin
~\citep{sander2024watermarking,golowich2024edit,hu2024inevitable}
These patterns remain imperceptible 
to humans yet detectable through algorithmic analysis, 
thereby promoting greater transparency and accountability in 
AI-generated content~\citep{GuoTSL0024,HouZHWCWSDKT24}.
Among various approaches, 
the green and red list-based watermarking 
is the most widely adopted~\citep{kirchenbauer2023watermark,zhao2023provable,lee-etal-2024-wrote,lu-etal-2024-entropy}. 
These approaches divide token candidates 
into two lists~(green and red) 
and bias the selection process toward tokens in the green list. 
Detection involves measuring the proportion of green list tokens 
in the generated output. 
While effective for natural language tasks, 
this approach does not easily apply to code generation, 
where structural and functional correctness are 
crucial~\citep{lee-etal-2024-wrote, hoang2024less}. 
Changes to syntax-critical tokens can cause 
compilation errors or alter program behavior.

We propose \texttt{STONE}~(\textbf{S}yntax \textbf{TO}ke\textbf{N} preserving cod\textbf{E} watermarking), 
a method that selectively embeds watermarks in non-syntax tokens. 
By excluding tokens that are critical to code execution, 
\texttt{STONE} reduces the risk of functional degradation 
while embedding robust, detectable patterns.
Additionally, we introduce \texttt{STEM}~(Summative Test metric for Evaluating code-waterMarking).
Existing studies on code watermarking rely on separate evaluation criteria,
using different metrics and measurement protocols across methods,
hindering fair and consistent comparison of performance.
\texttt{STEM} integrates correctness, detectability, and imperceptibility 
into a unified metric,
providing a fair basis for evaluating overall performance 
while allowing task-specific weighting.
We employ \texttt{STEM} to evaluate \texttt{STONE} 
and compare it with state-of-the-art watermarking methods.
Experiments across Python, C++, and Java 
show that \texttt{STONE} preserves functional correctness,
achieves balanced performance across all evaluation criteria,
and consistently outperforms prior methods in efficiency.

\begin{figure}[hbt!]
    \centering
        \includegraphics[width=\columnwidth]{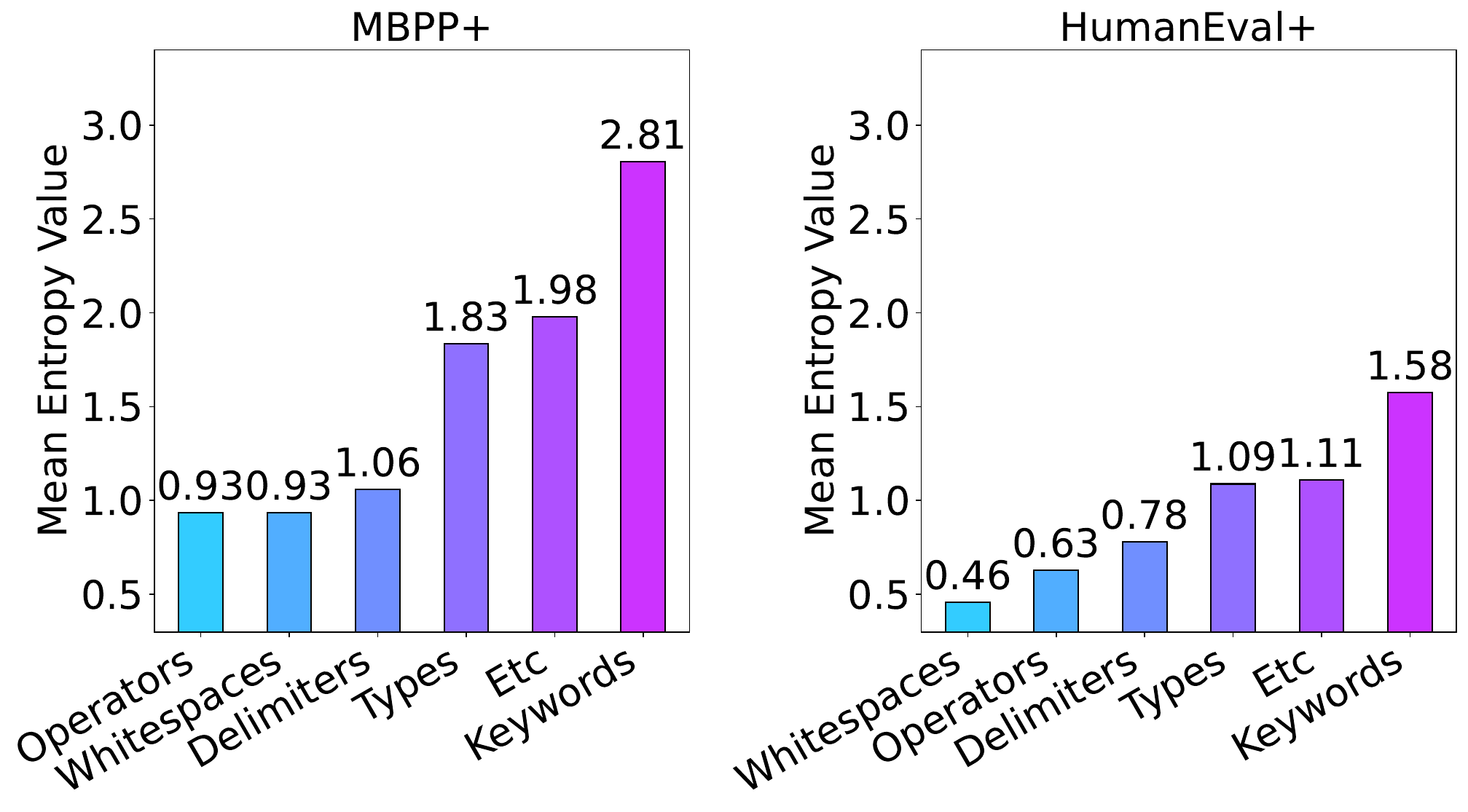}
    \caption{
    Entropy values by token type.
    Tokens that do not fall under keywords, whitespace, types, delimiters, or operators 
    are categorized as \emph{etc} tokens.
    The \emph{etc} category is the primary target for our proposed \texttt{STONE} 
    watermarking method.
    Token entropy is measured using Qwen2.5-Coder-7B.
    }\label{fig:preliminary_analysis}
\end{figure} 

\section{Motivation and Preliminary Analysis}
\label{sec:preliminary_analysis}

The recent code watermarking method SWEET~\citep{lee-etal-2024-wrote}
embeds watermarks in high-entropy tokens.
This design is motivated by the intuition that modifying tokens with higher
uncertainty is less likely to affect the perceived quality of the generated code.
In our preliminary analysis, we examine whether token entropy indeed varies
systematically across syntactic categories and how such variation relates to
potential impacts on code quality.

We consider two popular benchmarks, MBPP+ and HumanEval+, for Python code
generation and completion.
Each Python token is classified into one of five syntactic categories
(keywords, whitespace, types, delimiters, and operators), with all remaining
tokens grouped into an \emph{etc} category.
Details of the categorization procedure are provided in
Appendix~\ref{app:example_of_syntax_elements}.

We compute token entropy for each category using Shannon entropy,
following~\citet{lee-etal-2024-wrote}.
Given a token sequence \( y = (y_0, y_1, \ldots, y_N) \),
the entropy at generation step \( t \) is defined as:
\begin{equation}
H_t = - \sum_{i=1}^{|V|} P(y_t = v_i \mid y_{<t}) \log P(y_t = v_i \mid y_{<t}),
\end{equation}

where \( V \) denotes the vocabulary and \( y_{<t} \) the previously generated tokens.
This formulation captures the uncertainty of the model’s next-token prediction
at each generation step.

Figure~\ref{fig:preliminary_analysis} shows the mean entropy values for each
token category.
We observe that keywords exhibit the highest average entropy among all categories.
This can be attributed to their diverse roles across control flow, structural
definitions, and other syntactic constructs in Python.
In contrast, operators, delimiters, types, and whitespace follow more regular
and constrained syntactic patterns, resulting in comparatively lower entropy.

Although keywords have high entropy, watermarking them risks altering tokens that are essential to code syntax and semantics.
Even small perturbations to such tokens may lead to functional errors or
noticeable degradation in code quality.
In comparison, the \emph{etc} category contains tokens that are less critical
to core syntactic structure while still exhibiting relatively high entropy.
This observation suggests that embedding watermarks in non-syntactic tokens may preserve code quality better than approaches that rely solely on high-entropy tokens.

\section{Methodology}
\subsection{Syntax-Aware Watermarking: \texttt{STONE}}
\label{sec:approach}

\texttt{STONE} is a code watermarking method designed to preserve the quality of generated code.
This section introduces the watermark insertion and detection procedures of \texttt{STONE}.

\paragraph{Inserting Watermarks}
\label{sec:inserting_watermark}   

We treat keywords, whitespace, types, delimiters, and operators as
syntactic elements and embed watermarks only in non-syntactic tokens.
Algorithm~\ref{alg:zero_bit_generation} describes how \texttt{STONE} 
embeds watermarks during code generation.
Given a prompt $\mathbf{x}$ and previously generated tokens $\mathbf{y}_{<t}$, 
the language model $f_\text{LM}$ computes the initial probability distribution 
$\mathbf{p}_t$ at time step $t$.
\texttt{STONE} first samples a candidate token $\tilde{y}_t$ 
from the initial probability distribution. 
If the sampled token is not in the syntax element set, 
\texttt{STONE} then divides the vocabulary $\mathcal{V}$ into a green and red list 
according to the green token ratio $\gamma$.
It then increases the likelihood of sampling green-list tokens by adding a constant $\delta$ to their logit values.
Using the adjusted logits, 
\texttt{STONE} updates the probability 
distribution and samples the final token $y_t$ for time step $t$.

\begin{algorithm}[h!]\small
\caption{Insertion Algorithm of \texttt{STONE}}\label{alg:zero_bit_generation}
\begin{algorithmic}[1]
\Require Prompt $\mathbf{x}$,
         start token $y_0$,
         last generated tokens $\mathbf{y}_{<t} = (y_0, \dots, y_{t-1})$,
         vocabulary set $\mathcal{V}$,
         language model $f_{\text{LM}}$,
         syntax element set $S$,
         green list ratio $\gamma \in (0, 1)$,
         logit adjustment factor $\delta > 0$
\For{$t = 1, 2, 3, \dots$}
    \State Compute the logit vector:
    \[
    \mathbf{l}_t = f_{\text{LM}}(\mathbf{x}, \mathbf{y}_{<t})
    \]
    \State Compute the initial probability vector $\mathbf{p}_t$:
    \[
    p_{t,i} = \frac{e^{\mathbf{l}_t[i]}}{\sum_{j=1}^{|\mathcal{V}|} e^{\mathbf{l}_t[j]}},
    \quad \text{for } i \in \{1, \dots, |\mathcal{V}|\}
    \]
    \State Sample a candidate token $\tilde{y}_t \sim \mathbf{p}_t$
    \If{$\tilde{y}_t \notin S$}
        \State Compute a hash of token $y_{t-1}$ as a seed
        \State Split vocabulary indices into $\mathcal{G}_t$ and $\mathcal{R}_t$
        \State Increase logits for indices in $\mathcal{G}_t$ by $\delta$:
        \[
        \mathbf{l}_t[i] \leftarrow \mathbf{l}_t[i] + \delta,
        \quad \text{for } i \in \mathcal{G}_t
        \]
        \State Recompute the adjusted probability vector $\mathbf{p}'_t$:
        \[
        p'_{t,i} = \frac{e^{\mathbf{l}_t[i]}}{\sum_{j=1}^{|\mathcal{V}|} e^{\mathbf{l}_t[j]}},
        \quad \text{for } i \in \{1, \dots, |\mathcal{V}|\}
        \]
    \Else
        \State Set $\mathbf{p}'_t = \mathbf{p}_t$
    \EndIf
    \State Sample the final token $y_t \sim \mathbf{p}'_t$
\EndFor
\end{algorithmic}
\end{algorithm}

\paragraph{Detecting Watermarks}
\label{sec:detecting_watermark}

We assess how many green tokens are present in the code, 
and if this value surpasses a certain threshold, 
we conclude that the code was generated by an LLM. 
When identifying potential green tokens, we focus only on 
non-syntactic elements. 
Algorithm~\ref{alg:zero_bit_detection} 
details the watermark detection process.
$N^E$ represents the number of \emph{etc} tokens~(non-syntax tokens) in the code, 
and $N^E_G$ denotes the number of green tokens among them. 
We replicate the division of the vocabulary into green and red lists 
at each time step during code generation to determine whether 
a specific token is a green token.
If the $z$-score exceeds the predefined threshold $z_{\text{threshold}}$, 
the code is determined to be generated by an LLM.

\begin{algorithm}[hbt!]\small
\caption{Detection Algorithm of \texttt{STONE}}\label{alg:zero_bit_detection}
\begin{algorithmic}[1]
\Require Token sequence $\mathbf{X} = (X_0, \dots, X_{N-1})$,
         syntax element set $S$,
         green list ratio $\gamma \in (0, 1)$,
         $z$-score threshold $z_{\text{threshold}} > 0$
\State Initialize $N^E \gets 0$, $N^E_G \gets 0$
\For{$t = 1, 2, \dots, N - 1$}
    \If{$X_t \notin S$}
        \State $N^E \gets N^E + 1$
        \State Compute a hash of token $X_{t-1}$ as a seed
        \State Split vocabulary indices into $\mathcal{G}_t$ and $\mathcal{R}_t$
        \If{$X_t \in \mathcal{G}_t$}
            \State $N^E_G \gets N^E_G + 1$
        \EndIf
    \EndIf
\EndFor
\State Compute the $z$-score:
\[
z = \frac{N^E_G - \gamma N^E}{\sqrt{\gamma (1 - \gamma) N^E}}
\]
\If{$z > z_{\text{threshold}}$}
    \State \Return True \Comment{Sequence $\mathbf{X}$ is watermarked}
\Else
    \State \Return False \Comment{Sequence $\mathbf{X}$ is not watermarked}
\EndIf
\end{algorithmic}
\end{algorithm}

\subsection{Unified Evaluation Metric: \texttt{STEM}}
\label{sec:STEM}

Prior studies on code watermarking typically assess functional correctness, detectability, and imperceptibility independently~\citep{lee-etal-2024-wrote,yang2024srcmarker,li2024resilient,guan2024codeip}.
They adopt diverse definitions and evaluation protocols, which prevents consistent and fair comparison across methods.
For instance, a technique may improve detectability at the cost of correctness, leading to inconsistent evaluations.
\texttt{STEM} addresses this issue by unifying the three criteria into a single weighted metric $(\alpha, \beta, \zeta)$, where $\alpha+\beta+\zeta=1$, enabling fair and direct comparison across methods under configurable trade-offs.

\begin{equation}
\begin{aligned}
\text{\texttt{STEM}} = &\; \alpha \cdot \mathrm{Correctness}(C_{wm}) \\
&+ \beta \cdot \mathrm{Detectability}(C_{wm}, C_H) \\
&+ \zeta \cdot \mathrm{Imperceptibility}(C_{wm}, C).
\end{aligned}
\end{equation}

\paragraph{Correctness}\label{ssec:Correctness}
\emph{Correctness} is crucial for preventing watermarking from compromising code functionality.
Let \(C_{wm}\) denote the set of watermarked code samples.
We measure functionality preservation using the unbiased estimator of pass@k~\cite{chen2021evaluating}, which is standard for evaluating code generation correctness:
\begin{equation}
\mathrm{Correctness}(C_{wm}) =
\mathbb{E}_{C_{wm}}\!\left[ 1 - \frac{\binom{n-c}{k}}{\binom{n}{k}} \right],
\label{eq:correctness}
\end{equation}

where $n$ corresponds to the total number of generated solutions for a task, and
$c$ denotes the subset of those solutions that pass all test cases.
This estimator measures the probability that at least one of the $k$ generated solutions is functionally correct.
In the context of watermarking, it captures whether watermark insertion
introduces execution failures or alters program semantics.
A higher correctness score indicates that watermark insertion does not hinder program execution.

\paragraph{Detectability}\label{ssec:Detectability}

\emph{Detectability} measures how effectively a watermark can be identified in generated code.
We assess detectability using a $z$-score that quantifies the proportion of green-list tokens,
and aggregate scores by computing the AUROC between human-written and watermarked corpora.

For a generated sequence \(X_{wm} = (X_{0}, X_{1}, \dots, X_{N-1})\),
the $z$-score of green-token ratio is defined as
\begin{equation}
z(X_{wm}) =
\frac{|X_{wm}|_{G} - \gamma |X_{wm}|}
     {\sqrt{\gamma (1-\gamma) |X_{wm}|}},
\label{eq:zscore}
\end{equation}

where \(|X_{wm}|_{G}\) is the number of green tokens in \(X_{wm}\)
and \(\gamma\) is the expected green-token ratio in non-watermarked code.
Higher \(z\) values indicate stronger evidence that the code is watermarked.
The $z$-score serves as a continuous detection statistic, 
enabling threshold-based discrimination between watermarked and non-watermarked sequences.
By varying the threshold $\tau$, different operating points in terms of true and false positive rates are obtained, forming the ROC curve.

Given a corpus of human-written code \(C_H = \{X^{(i)}_H\}\)
and watermarked code \(C_{wm} = \{X^{(j)}_{wm}\}\),
the true positive rate~(TPR) and false positive rate~(FPR) at threshold \(\tau\)
are defined as follows:
\begin{align}
\text{TPR}(\tau) &=
\frac{\sum_{j=1}^{J} \mathbf{1}\!\bigl[z(X_{wm}^{(j)}) > \tau\bigr]}{J}. \label{eq:tpr}\\
\text{FPR}(\tau) &=
\frac{\sum_{i=1}^{I} \mathbf{1}\!\bigl[z(X_{H}^{(i)}) > \tau\bigr]}{I}. 
\label{eq:fpr}
\end{align}
The area under the ROC curve (AUROC) is then
\begin{multline}\label{eq:detectability}
\mathrm{Detectability}(C_{wm}, C_H) = \\
\int_{0}^{1} \text{TPR}(\text{FPR})\, d(\text{FPR}),
\end{multline}

where higher AUROC values indicate stronger separability between
watermarked and human-written code.

\paragraph{Imperceptibility}\label{ssec:imperceptibility}
\emph{Imperceptibility} concerns the preservation of token probability 
distributions consistent with normal model outputs~\citep{huang-wan-2025-waterpool}. 
Since code LLMs follow stable and highly structured token distributions, 
distortions introduced by watermarking can form statistical signals that are exploitable by adversaries without access to the secret key.
Accordingly, following perplexity-based approaches in text watermarking~\citep{kirchenbauer2023watermark} 
and studies on code generation~\citep{magnusson2024paloma}, 
we define imperceptibility as the deviation in token probability 
distributions introduced by watermarking, measured through perplexity shifts 
computed with code LLMs.

Let \(C_{wm}\) and \(C\) denote the corpora of watermarked and non-watermarked code, respectively.
Perplexity is computed over the watermarked corpus
\({C}_{wm}=\{X_{wm}^{(1)},\dots,X_{wm}^{(J)}\}\) as:
\begin{multline}\label{eq:ppl}
\operatorname{PPL}(C_{wm}) = \\
\frac{1}{\lvert C_{wm}\rvert}
\sum_{j=1}^{\lvert C_{wm}\rvert}
\exp\!\Bigl(
-\mathbb{E}_{i}\!
\left[
\log P\!\left(y_i^{(j)} \mid y_{<i}^{(j)}\right)
\right]
\Bigr),
\end{multline}

where $P(y_i^{(j)}|y_{<i}^{(j)})$ represents the probability of token $y_i^{(j)}$ 
given its preceding context. 
An effectively concealed watermark minimizes perplexity shifts, 
ensuring imperceptibility.
We define the imperceptibility metric as:
\begin{multline}\label{eq:imperceptibility}
\mathrm{Imperceptibility}(C_{wm}, C) = \\
1 - \frac{\bigl|\operatorname{PPL}(C_{wm}) - \operatorname{PPL}(C)\bigr|}
         {\operatorname{PPL}(C)}.
\end{multline}

This metric quantifies the perplexity-based difference between 
LLM-generated code with watermark~(\(C_{wm}\)) and non-watermarked code~(\(C\)).
It computes the relative difference by normalizing the absolute gap 
by the perplexity of $C$, 
providing a scale-invariant measure of distributional shift.
Higher values indicate minimal impact on token distributions and thus stronger imperceptibility,
while lower or negative values reflect greater disruption due to watermark insertion.

\section{Experimental Setup}
\label{sec:experimental_setup}

\paragraph{Datasets.}
For our experiments, we evaluate our approach across three programming languages:
Python, C++, and Java.
We use four benchmark datasets: HumanEval+ and MBPP+~\citep{evalplus} for Python,
and HumanEvalPack~\citep{MuennighoffLZZH24} for C++ and Java.
Table~\ref{tab:data_statistics} reports token-length statistics of the evaluation datasets,
based on the Qwen2.5-Coder-7B tokenizer. 
The datasets span from short snippets to long sequences, allowing evaluation under varying code lengths and structural complexity.

\begin{table}[h!]
\centering
\small
\captionsetup{skip=6pt}
\setlength{\tabcolsep}{3.2pt}
\renewcommand{\arraystretch}{1.15}
\resizebox{\columnwidth}{!}{%
\begin{tabular}{lrrrrrr}
\toprule
Dataset & Problems & Test Cases & Max & Min & Mean & Std. \\
\midrule
MBPP+      & 399 & 105 & 341 & 11 &  40.25 &  34.91 \\
HumanEval+ & 164 & 764 & 558 & 43 & 188.28 &  87.32 \\
HEP-C++    & 164 & 764 & 697 & 42 & 223.10 & 107.57 \\
HEP-Java   & 164 & 764 & 660 & 56 & 237.36 & 105.09 \\
\bottomrule
\end{tabular}%
}
\caption{Token length statistics of solution code across evaluation datasets, along with the total number of problems and the average number of test cases per problem.}
\label{tab:data_statistics}
\end{table}

\paragraph{Baselines.}
We consider the following three watermarking approaches for comparison.
(1) KGW~\citep{kirchenbauer2023watermark} generates text by dividing the vocabulary into a green and red list 
at each time step of token generation, 
increasing the probability of generating tokens from the green list. 
(2) EWD~\citep{lu-etal-2024-entropy} leverages entropy-based token weighting during detection, 
giving more influence to high-entropy tokens, 
thus enhancing detection in texts with varying entropy distributions.
(3) SWEET~\citep{lee-etal-2024-wrote} selectively embeds watermarks in high-entropy tokens, 
addressing detection challenges posed by the low entropy in code.
We exclude CodeIP~\citep{guan2024codeip} 
from our baselines as it requires additional training for watermark insertion, 
while our approach and the selected baselines are training-free methods.
As part of our comparative analysis, 
we reproduce CodeIP and report its 
experimental results in Appendix~\ref{app:codeip}.

\paragraph{Base Model.} We use Qwen2.5-Coder-7B~\citep{hui2024qwen2} for our experiments.
We also report additional results with Llama-3.1-8B 
in Appendix~\ref{app:llama3.1}, 
which further demonstrate the generalizability of our approach 
across different model architectures.

\paragraph{Evaluation Metric.} 
We evaluate watermarking methods across three dimensions: 
Correctness, Detectability, and Imperceptibility. 
Correctness uses pass@k~(k=1,5), 
detectability is measured by AUROC with a $z$-score test, 
and imperceptibility by perplexity computed with StarCoder2-7B. 
We use the \texttt{STEM} metric to integrate these three axes into a single composite score.
For the main results (Tables~\ref{tab:experimental_results_main_a}
and~\ref{tab:experimental_results_main_b}),
we adopt an equal-weight configuration $(\tfrac13,\tfrac13,\tfrac13)$.
We further analyze weight settings with alternative configurations in Appendix~\ref{app:stem-grid}.

\paragraph{Implementation Details.}

During code generation, 
we apply top-$k$ sampling with $k=50$, 
restricting the candidate token pool to the top 50 tokens with 
the highest logits and 
normalizing their probabilities 
before sampling. 
The temperature, which controls the sharpness of 
the token distribution, is fixed at 1.0. 
We set \(\gamma = 0.5\) and \(\delta = 1.0\) 
for the MBPP+ dataset, 
while for the HumanEval+ and HumanEvalPack datasets (C++ and Java),
we set \(\gamma = 0.5\) and \(\delta = 0.5\).
We conduct all experiments on an NVIDIA A6000 GPU.

\begin{table*}[hbt!]
\centering
\captionsetup{skip=6pt}
\footnotesize
\setlength{\tabcolsep}{3.2pt}
\renewcommand{\arraystretch}{1.15}
\begin{tabular}{@{}lcccccccc@{}}
\noalign{\hrule height 0.8pt}
\multirow{2}{*}{Method} &
\multicolumn{4}{c}{MBPP+} &
\multicolumn{4}{c}{HumanEval+} \\
\cmidrule(lr){2-5}\cmidrule(lr){6-9}
& Correctness & Detectability & Imperceptibility & \texttt{STEM}       &
  Correctness & Detectability & Imperceptibility & \texttt{STEM}  \\
\midrule
KGW   & 0.499 & 0.831 & \textbf{0.994} & 0.775
      & 0.573 & 0.523 & \textbf{0.986} & 0.694 \\
EWD   & 0.499 & 0.965 & \textbf{0.994} & 0.819
      & 0.573 & 0.730 & \textbf{0.986} & 0.763 \\
SWEET & 0.502 & 0.867 & 0.992 & 0.787
      & 0.574 & 0.710 & 0.978 & 0.754 \\
\texttt{STONE} 
      & \textbf{0.571} & \textbf{0.982} & 0.990 & \textbf{0.848}
      & \textbf{0.587} & \textbf{0.777} & 0.978 & \textbf{0.781} \\
\noalign{\hrule height 0.8pt}
\end{tabular}
\caption{Experimental results on MBPP+ and HumanEval+. Metrics include correctness, detectability, imperceptibility, and the integrated \texttt{STEM} score under the equal-weight configuration.}
\label{tab:experimental_results_main_a}
\end{table*}

\begin{table*}[hbt!]
\centering
\captionsetup{skip=6pt}
\footnotesize
\setlength{\tabcolsep}{3.2pt}
\renewcommand{\arraystretch}{1.15}
\begin{tabular}{@{}lcccccccc@{}}
\noalign{\hrule height 0.8pt}
\multirow{2}{*}{Method} &
\multicolumn{4}{c}{HEP-C++} &
\multicolumn{4}{c}{HEP-Java} \\
\cmidrule(lr){2-5}\cmidrule(lr){6-9}
& Correctness & Detectability & Imperceptibility & \texttt{STEM} &
  Correctness & Detectability & Imperceptibility & \texttt{STEM} \\
\midrule
KGW   & 0.576 & 0.621 & \textbf{0.993} & 0.730
      & 0.387 & 0.546 & \textbf{0.993} & 0.642 \\
EWD   & 0.576 & 0.681 & \textbf{0.993} & 0.750
      & 0.387 & 0.646 & \textbf{0.993} & 0.675 \\
SWEET & 0.584 & 0.641 & 0.979 & 0.735
      & 0.413 & 0.580 & 0.901 & 0.631 \\
\texttt{STONE}
      & \textbf{0.622} & \textbf{0.729} & 0.990 & \textbf{0.780}
      & \textbf{0.445} & \textbf{0.721} & 0.979 & \textbf{0.715} \\
\noalign{\hrule height 0.8pt}
\end{tabular}
\caption{Experimental results on HumanEvalPack-C++ and HumanEvalPack-Java. Metrics include correctness, detectability, imperceptibility, and the integrated \texttt{STEM} score under the equal-weight configuration.}
\label{tab:experimental_results_main_b}
\end{table*}

\section{Experiments and Results}

We pose three research questions and provide a thorough analysis of each:
\begin{itemize}
\item \textbf{RQ1:} Can \texttt{STONE} preserve functional correctness?
\item \textbf{RQ2:} Does \texttt{STONE} achieve balanced performance across correctness, detectability, and imperceptibility as measured by \texttt{STEM}?
\item \textbf{RQ3:} How efficient is \texttt{STONE} in watermark insertion and detection?
\end{itemize}
We address these questions through three analytical dimensions:
(1) \textbf{Functionality Preservation}, which verifies that watermark insertion does not alter program behavior;
(2) \textbf{Balanced Performance under \texttt{STEM}}, which examines whether \texttt{STONE} maintains a well-balanced trade-off among correctness, detectability, and imperceptibility; and
(3) \textbf{Efficiency}, which measures the computational overhead of watermark insertion and detection.
\paragraph{RQ1}
We examine whether \texttt{STONE} embeds watermarks without 
affecting the functional correctness of generated code.
Tables~\ref{tab:experimental_results_main_a} and \ref{tab:experimental_results_main_b} show that \texttt{STONE} 
improves correctness by 7.57\% over SWEET on average across all benchmarks.
The token-level analysis in 
Appendix~\ref{app:sweet_watermark_token_selection} 
indicates that the entropy-based token selection method used by SWEET 
often chooses syntax-related elements—such as delimiters, whitespace, 
keywords, types, and operators—that are essential for correct execution.
For example, modifying delimiters like colons~(`:') or brackets~(`{}', `[]') 
can cause parsing errors, and changing operators or keywords 
(e.g., replacing `+' with `-' or `True' with `False') 
can alter program logic or control flow.
Even under the optimal configuration, syntax tokens account for 12.6\% of 
the tokens selected for watermarking in the HumanEval+ dataset, 
with delimiters and whitespace forming the majority.
This inclusion of syntax elements explains the observed reduction in correctness for SWEET.
In contrast, \texttt{STONE} excludes syntax tokens 
and embeds watermarks only into non-syntactic elements, 
which prevents structural and logical errors during execution 
and allows the generated code to preserve its functional behavior.

\paragraph{RQ2} 
We examine whether \texttt{STONE} achieves balanced performance across correctness, detectability, and imperceptibility.
Baselines such as KGW and EWD watermark every token, which dilutes the watermark signal
and lowers AUROC scores.
SWEET modifies high-entropy tokens, including syntax elements, improving detectability
but reducing syntactic stability.
In contrast, \texttt{STONE} embeds watermarks only in non-syntax tokens, preserving syntactic integrity while keeping the signal distinct.
This selective design enables \texttt{STONE} to achieve higher detection accuracy without compromising code imperceptibility.

Imperceptibility results reinforce this finding.
KGW and EWD maintain reasonable imperceptibility by uniformly adjusting token probabilities, but their adjustments can still introduce subtle artifacts.
SWEET often produces more noticeable distributional shifts because it alters syntax tokens.
\texttt{STONE}, in contrast, maintains low perplexity differences by leaving syntax untouched and modifying only non-syntax tokens.
As a result, \texttt{STONE} maintains imperceptibility without sacrificing
correctness or detection robustness.
We quantify this balanced behavior using the \texttt{STEM} metric, which under the equal-weight configuration provides a neutral evaluation across correctness, detectability, and imperceptibility.
Unlike other watermarking methods that exhibit clear trade-offs among these criteria, \texttt{STONE} achieves consistently strong overall performance.
A comprehensive grid analysis across 66 weight settings~(Appendix~\ref{app:stem-grid}) further confirms that this advantage remains stable and does not depend on any specific weighting scheme.

\begin{table*}[hbt!]
\centering
\small
\setlength{\tabcolsep}{3.5pt}
\renewcommand{\arraystretch}{1.15}
\begin{tabular}{lcccccccc}
\noalign{\hrule height 0.8pt}
\multirow{2}{*}{Method} &
\multicolumn{2}{c}{MBPP+} &
\multicolumn{2}{c}{HumanEval+} &
\multicolumn{2}{c}{HEP-C++} &
\multicolumn{2}{c}{HEP-Java} \\
\cmidrule(lr){2-3}\cmidrule(lr){4-5}\cmidrule(lr){6-7}\cmidrule(lr){8-9}
& Insertion & Detection &
  Insertion & Detection &
  Insertion & Detection &
  Insertion & Detection \\
\midrule
KGW
& 3320 & 12.90
& 1268 & 4.19
& 1308 & 8.01
& 1506 & 8.95 \\
EWD
& 3320 & 100.43
& 1268 & 34.51
& 1308 & 56.85
& 1506 & 59.84 \\
SWEET
& 3300 & 100.94
& 1270 & 34.68
& 1454 & 58.08
& 1480 & 59.22 \\
\texttt{STONE}
& 3266 & 13.27
& 1277 & 4.62
& 1300 & 8.48
& 1459 & 9.24 \\
\noalign{\hrule height 0.8pt}
\end{tabular}
\caption{Comparison of total insertion and detection time (measured in seconds) across different watermarking methods. The reported measurement is performed at the dataset level.}
\label{tab:generation_detection_time}
\end{table*}

\paragraph{RQ3}
We analyze the computational overhead of watermarking in terms of insertion and detection time.
Insertion time denotes the duration required to generate watermarked code, whereas detection time measures the time needed to verify a watermark.
Table~\ref{tab:generation_detection_time} summarizes the results for each method.
All approaches exhibit comparable insertion times, but \texttt{STONE} achieves detection speeds that are on average 86\% faster than those of SWEET and EWD
at the dataset level.
This improvement results from differences in detection mechanisms.
SWEET and EWD employ entropy-based approaches that compute token probability distributions, which substantially increase computational overhead.
In contrast, \texttt{STONE} verifies watermarks using a predefined green list of non-syntax tokens, reducing detection time while maintaining reliable detectability.

\section{Analysis}

\subsection{Trade-off Between Code Quality and Detection Performance}
\label{sec:tradeoff}

\begin{figure}[hbt!]
    \centering
        \includegraphics[width=8cm]{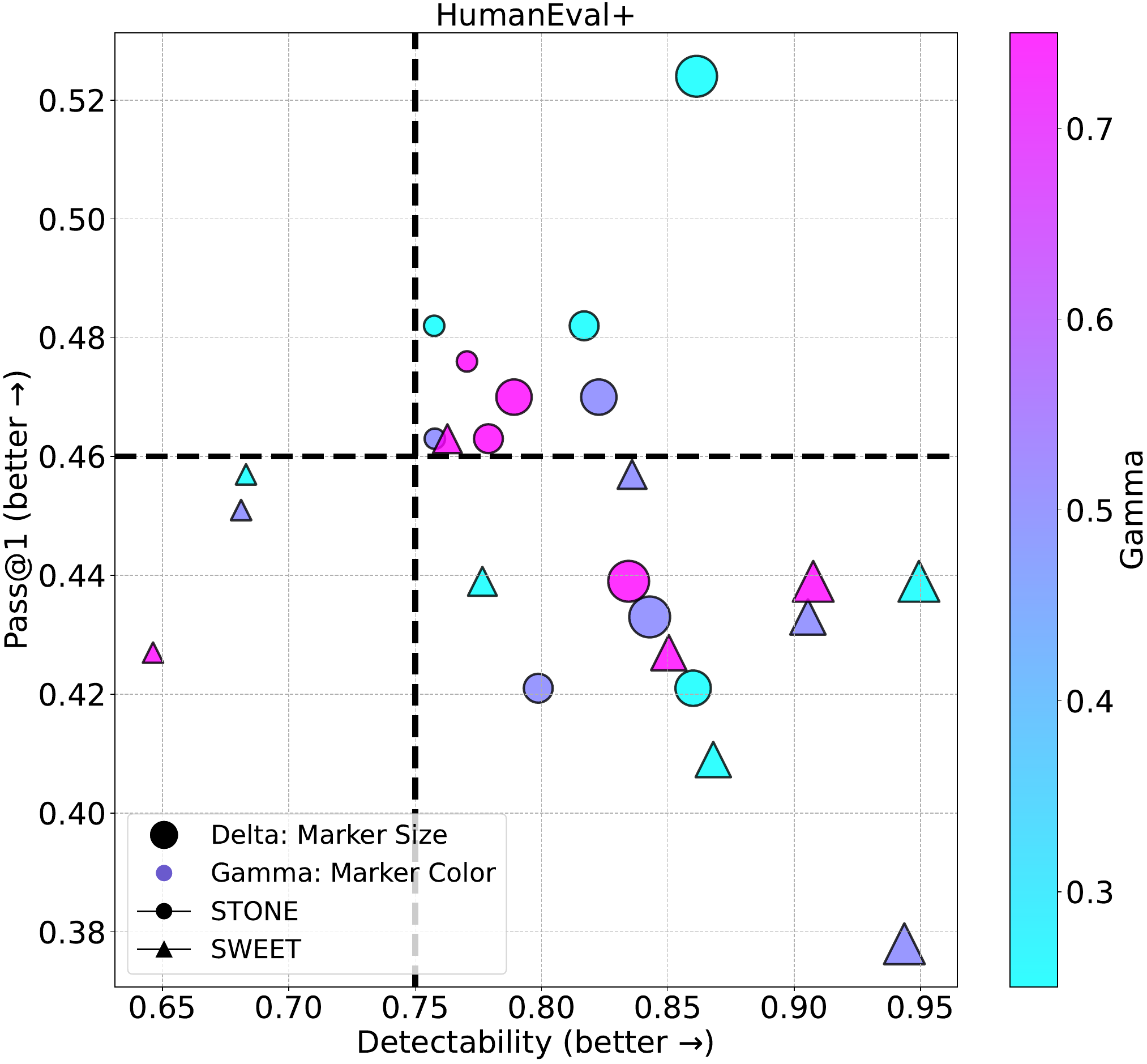}
    \caption{
    The trade-off between pass@1~(Y-axis) and detectability~(X-axis). 
    We analyze the impact of the green list ratio $\gamma$~(indicated by marker color) and 
    the watermark strength $\delta$~(indicated by marker size). 
    Circular and triangular markers represent \texttt{STONE} and SWEET, respectively.
    }\label{fig:delta_gamma}
\end{figure}

We analyze the trade-off between code quality and watermark detectability 
as influenced by two watermarking parameters. 
The parameter \(\gamma\) controls the proportion of the vocabulary designated 
as the green list, 
while \(\delta\) determines the extent to which the selection probability of 
green tokens is amplified.
We conduct an analysis on the HumanEval+ dataset
by varying the values of \(\delta\) and \(\gamma\).
Figure~\ref{fig:delta_gamma} shows the results in a bubble chart, 
revealing a clear trade-off between code quality and detectability.
The circular markers represent the 
pass@1 and detection performance~(AUROC) of \texttt{STONE}, 
while the triangular markers indicate those of SWEET. 
Among the four regions divided by the two black dashed lines,
the upper-right quadrant corresponds to configurations where both
code quality and detection performance are high, indicating a desirable
balance between utility and detectability. 
We observe that, while only a single configuration of 
SWEET falls into this balanced region, 
multiple configurations of \texttt{STONE}—spanning various combinations of 
\(\gamma\) and \(\delta\)—lie within it. 
This indicates that \texttt{STONE} achieves a more balanced trade-off between
functionality and watermark detectability compared to SWEET.

\subsection{Robustness under Adversarial Attacks}
\label{sec:robustness}

\begin{figure}[hbt!]
    \centering
        \includegraphics[width=7.5cm]{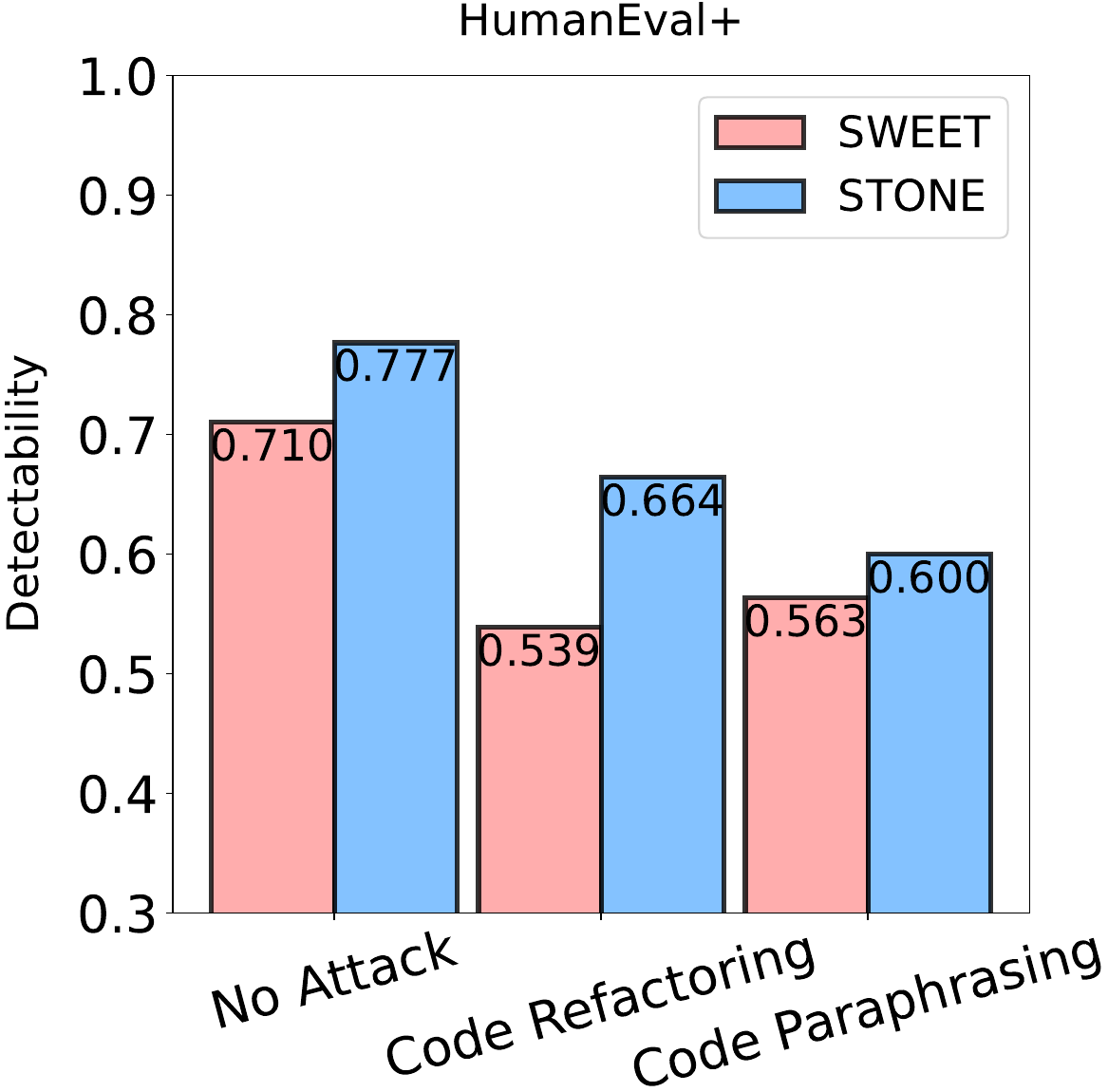}
    \caption{
    We evaluate robustness by applying code refactoring and paraphrasing attacks
to watermarked code and measuring detection performance.
Results compare the resilience of SWEET and \texttt{STONE} under both attacks. 
    }\label{fig:attack_analysis}
\end{figure}

Although a watermark is embedded in the code, 
a malicious user may remove it by modifying the code. 
Therefore, the watermark must remain robust
against attacks like code modifications.
We assess the resilience of watermarks of \texttt{STONE}  
against code alterations by applying two types of 
attacks: (1) a code refactoring attack using a commercial service\footnote{\url{https://codepal.ai/code-refactor}} and 
(2) a code paraphrasing attack using ChatGPT~(GPT-4o)\footnote{Prompt: Please paraphrase the following code.}. 
We conduct this analysis on the HumanEval+ dataset.
Figure~\ref{fig:attack_analysis} 
shows the classification performance (i.e., detectability)
in distinguishing between 
watermarked LLM-generated code and human-written code after these attacks 
are applied to code watermarked by \texttt{STONE} and SWEET.
The experimental results show that \texttt{STONE} is more robust 
than SWEET against both types of attacks. 
Refactoring reorganizes code while keeping syntax boundaries, 
so \texttt{STONE} preserves many watermarks inserted into non-syntax tokens. 
Paraphrasing changes variable names and expressions, which directly correspond to the embedding targets of \texttt{STONE}, 
and thus decreases detectability. 
SWEET shows the opposite tendency because refactoring alters its watermarking targets.
\texttt{STONE} maintains higher robustness under both attacks by embedding
watermarks into a syntax-filtered token space that remains stable under
structural and lexical modifications.
This result confirms that \texttt{STONE} maintains consistent detection robustness across different attack types.

\section{Related Work}
Post-hoc detection methods~\citep{Xu_Sheng_2024,betweenlineofcode24}
identify machine-generated code after generation 
using statistical cues such as perplexity or syntax differences.
In contrast, watermarking embeds provenance signals 
during generation to enable direct identification~\citep{kirchenbauer2023watermark, guan2024codeip,park2025watermod}.
Our work focuses on the latter approach, 
developing a syntax-aware watermarking method for code generation.

Among watermarking approaches, \citet{lee-etal-2024-wrote} proposed SWEET,
a method that selectively embeds watermarks in high-entropy tokens.
While SWEET demonstrated improvements in detection capability 
and code quality, our preliminary analysis revealed that approximately 
12.6\% of these high-entropy tokens correspond to syntax elements 
such as reserved keywords. 
Modifying these critical tokens can disrupt program logic, 
resulting in degraded functional correctness.
\citet{guan2024codeip} introduced CodeIP, a grammar-guided 
watermarking approach by utilizing type prediction to maintain 
syntactic validity. 
This method ensured that the generated code remained grammatically correct. 
However, the reliance on type prediction requires significant 
computational resources, limiting its practicality for 
large-scale code generation tasks.
In contrast to these approaches, our method explicitly excludes
syntax-critical tokens from watermark insertion.
Our approach ensures that essential structural elements, 
like keywords and operators, remain unchanged, 
thereby preserving functional correctness without additional 
computational overhead.
ACW~\citep{li2024resilient} and SrcMarker~\citep{yang2024srcmarker} 
employed watermarking techniques based on code transformations, 
such as semantic-preserving edits and variable renaming. 
While these approaches preserved program functionality, 
the resulting modifications often introduce unnatural code patterns, 
which can increase the risk of watermark removal by adversaries.

\section{Conclusion}
\label{sec:conclusion}
We present \texttt{STONE}, 
a syntax-aware watermarking method
that preserves functional correctness 
by excluding syntax-critical tokens
during watermark insertion. 
We also propose \texttt{STEM}, 
a unified evaluation metric integrating correctness, detectability, and
imperceptibility 
for balanced assessment. 
Experiments on Python,
C++, and Java show that \texttt{STONE} outperforms
baseline methods without degrading code quality and reduces watermark
detection time compared with entropy-based approaches. 
Evaluation under the \texttt{STEM} metric 
demonstrates balanced performance across weight 
configurations, and further analysis confirms that 
\texttt{STONE} maintains stable and
robust performance under code modifications. 

\section*{Limitations}
\label{sec:limitations}

While \texttt{STONE} provides a robust approach to functionality-preserving
watermarking, it has several limitations.

First, the watermarking capacity is based on the density of 
non-syntactic tokens. 
The conventional coding styles in our benchmark datasets 
provided ample opportunities for watermark insertion. 
This efficacy, however, can be compromised in unconventional code, 
such as obfuscated scripts 
or solutions for code golf, 
where such tokens are sparse. 
Under these conditions, the embedded watermark may prove too sparse 
for reliable detection.

Second, although our evaluation demonstrates resilience 
to general attacks 
like paraphrasing and refactoring, 
a more sophisticated threat model considers an adversary with specific 
knowledge of the \texttt{STONE} algorithm. 
Such an adversary may directly attack non-syntactic tokens 
by systematically renaming all variables and removing comments. 
This class of attack poses a more significant threat to watermark 
integrity than general modifications. 
Developing countermeasures against such targeted, 
algorithm-aware attacks 
remains an important direction for future research.

\section*{Acknowledgments}
This research was supported by the NRF grant~(RS-2025-00562134) 
and the AI Graduate School Program~(RS-2020-II201361) funded by the Korean government.

\bibliography{custom}

\appendix

\section{Performance Analysis of CodeIP}
\label{app:codeip}

\begin{table}[h!]
\centering
\footnotesize
\setlength{\tabcolsep}{2.5pt}
\renewcommand{\arraystretch}{1.12}
\resizebox{\columnwidth}{!}{%
\begin{tabular}{@{}lcccc@{}}
\noalign{\hrule height 0.8pt}
Dataset & Correct. & Detect. & Imperc. & PPL (NW$\to$WM) \\
\midrule
MBPP+       & 0.093 & 0.962 & -0.246 & (3.504$\to$7.869) \\
HumanEval+  & 0.018 & 0.945 & -0.075 & (3.276$\to$6.798) \\
HEP-C++     & 0.000 & 0.994 & -0.628 & (2.621$\to$6.890) \\
HEP-Java    & 0.073 & 0.994 & -1.013 & (2.426$\to$7.310) \\
\noalign{\hrule height 0.8pt}
\end{tabular}%
}
\caption{Evaluation of CodeIP on four datasets. Correctness: pass@1; Detectability: AUROC; Imperceptibility: PPL change from non-watermarked to watermarked code.}
\label{tab:codeip}
\end{table}

Table~\ref{tab:codeip} reports the reproduced results for CodeIP,
a grammar-guided multi-bit watermarking method that employs a type prediction model trained on CodeSearchNet for grammar-aware token selection during code generation.
All experiments are based on the original implementation and evaluated on four benchmarks: MBPP+, HumanEval+, HumanEvalPack-C++, and HumanEvalPack-Java.

CodeIP maintains high detectability, achieving AUROC values above 0.94 across all datasets.
This result indicates that grammar-based watermark insertion enables reliable message extraction.
However, this detection capability is accompanied by substantial drops in correctness and imperceptibility.
The pass@1 scores decrease sharply, approaching zero on HumanEval+ and HumanEvalPack-C++, which shows that syntactic validity enforced by the type predictor does not necessarily imply functional correctness.
The large perplexity gap of up to +4.9 indicates that the interaction between watermarking and type-prediction constraints can shift the token distribution, leading to deviations from the model’s natural generation pattern.

\begin{table}[h!]
\centering
\footnotesize
\setlength{\tabcolsep}{2.5pt}
\renewcommand{\arraystretch}{1.12}
\resizebox{\columnwidth}{!}{%
\begin{tabular}{@{}lcccc@{}}
\noalign{\hrule height 0.8pt}
Weight $(\alpha,\beta,\zeta)$ & MBPP+ & HumanEval+ & HEP-C++ & HEP-Java \\
\midrule
$(\tfrac{1}{3},\,\tfrac{1}{3},\,\tfrac{1}{3})$ & 0.270 & 0.296 & 0.122 & 0.018 \\
$(0.5,\,0.25,\,0.25)$                         & 0.241 & 0.274 & 0.095 & 0.041 \\
$(0.25,\,0.5,\,0.25)$                         & 0.381 & 0.361 & 0.267 & 0.145 \\
$(0.25,\,0.25,\,0.5)$                         & 0.325 & 0.331 & 0.174 & 0.084 \\
\noalign{\hrule height 0.8pt}
\end{tabular}%
}
\caption{\texttt{STEM} scores of CodeIP under different $(\alpha,\beta,\zeta)$ on MBPP+, HumanEval+, HumanEvalPack-C++, and HumanEvalPack-Java.}
\label{tab:codeip_stem}
\end{table}

Table~\ref{tab:codeip_stem} presents the \texttt{STEM} scores of CodeIP under different weight settings.
Across all configurations, detectability remains consistently high, while correctness- or imperceptibility-focused \texttt{STEM} scores are significantly lower than those of other methods.
Even under weight settings emphasizing correctness or imperceptibility, CodeIP does not achieve balanced performance, showing limited robustness under varying evaluation preferences.
These findings demonstrate that maximizing watermark detectability alone severely degrades code quality.
In contrast, \texttt{STONE} embeds watermarks only within non-syntax tokens and achieves stable performance across correctness, detectability, and imperceptibility as captured by the \texttt{STEM} metric.

\section{Experimental Results on Llama-3.1-8B}
\label{app:llama3.1}

Tables~\ref{tab:llama8b_a} and~\ref{tab:llama8b_b} summarize the experimental results obtained with the Llama-3.1-8B model on MBPP+ and HumanEval+.
\texttt{STONE} achieves the highest \texttt{STEM} scores across all datasets and weight settings, demonstrating that the non-syntax watermarking strategy generalizes well to larger models.
EWD shows slightly higher detectability on HumanEval+, which may be related to
differences in the output distribution of Llama-3.1-8B under uniform insertion.
In this setting, the cumulative bias introduced by EWD is more effectively reinforced across long and diverse sequences, leading to a small gain in detectability.
Such an effect is not observed under Qwen2.5-Coder-7B, whose flatter distributions yield comparable results between EWD and SWEET.
Overall, \texttt{STONE} preserves stronger correctness and achieves a more balanced trade-off, resulting in the highest overall \texttt{STEM} scores.

Across different weight configurations, \texttt{STONE} maintains stable performance even when correctness or imperceptibility is prioritized.
The method remains competitive under detectability-oriented settings without functional degradation.
This consistency confirms that syntax-aware watermarking effectively preserves code quality while sustaining robust watermark signals.

\begin{table}[h!]
\centering
\small
\setlength{\tabcolsep}{4pt}
\renewcommand{\arraystretch}{1.15}
\begin{adjustbox}{max width=\columnwidth}
\begin{tabular}{clcccccc}
\toprule
\multicolumn{2}{c}{\multirow{2}{*}{Method}} &
\multicolumn{3}{c}{MBPP+} &
\multicolumn{3}{c}{HumanEval+} \\
\cmidrule(lr){3-5}\cmidrule(lr){6-8}
\multicolumn{2}{c}{} &
\multicolumn{1}{c}{Correct.} & \multicolumn{1}{c}{Detect.} & \multicolumn{1}{c}{Imperc.} &
\multicolumn{1}{c}{Correct.} & \multicolumn{1}{c}{Detect.} & \multicolumn{1}{c}{Imperc.} \\
\midrule
\multirow{4}{*}{}
& \raggedleft KGW   & 0.414 & 0.831 & 0.953 & 0.323 & 0.507 & \textbf{0.989} \\
& EWD               & 0.414 & 0.931 & 0.953 & 0.323 & \textbf{0.755} & \textbf{0.989} \\
& SWEET             & 0.404 & 0.927 & \textbf{0.968} & 0.354 & 0.711 & 0.978 \\
& \texttt{STONE}    & \textbf{0.446} & \textbf{0.945} & 0.954 & \textbf{0.372} & 0.741 & 0.983 \\
\bottomrule
\end{tabular}
\end{adjustbox}
\caption{Experimental results on MBPP+ and HumanEval+ (Llama-3.1-8B): Correctness, Detectability, and Imperceptibility.}
\label{tab:llama8b_a}
\end{table}

\begin{table}[h!]
\centering
\small
\setlength{\tabcolsep}{3pt}
\renewcommand{\arraystretch}{1.15}
\begin{adjustbox}{max width=\columnwidth}
\begin{tabular}{@{}lcccccccc@{}}
\noalign{\hrule height 0.8pt}
\multirow{2}{*}{Weight $(\alpha,\beta,\zeta)$} &
\multicolumn{4}{c}{MBPP+} &
\multicolumn{4}{c}{HumanEval+} \\
\cmidrule(lr){2-5}\cmidrule(lr){6-9}
& KGW & EWD & SWEET & \texttt{STONE} 
& KGW & EWD & SWEET & \texttt{STONE} \\
\midrule
$(\tfrac{1}{3},\,\tfrac{1}{3},\,\tfrac{1}{3})$
& 0.733 & 0.766 & 0.766 & \textbf{0.782}
& 0.606 & 0.689 & 0.681 & \textbf{0.699} \\
$(0.5,\,0.25,\,0.25)$
& 0.653 & 0.678 & 0.676 & \textbf{0.698}
& 0.535 & 0.598 & 0.599 & \textbf{0.617} \\
$(0.25,\,0.5,\,0.25)$
& 0.757 & 0.807 & 0.806 & \textbf{0.823}
& 0.582 & 0.706 & 0.688 & \textbf{0.709} \\
$(0.25,\,0.25,\,0.5)$
& 0.788 & 0.813 & 0.817 & \textbf{0.825}
& 0.702 & 0.764 & 0.755 & \textbf{0.770} \\
\noalign{\hrule height 0.8pt}
\end{tabular}
\end{adjustbox}
\caption{
\texttt{STEM} scores on MBPP+ and HumanEval+ (Llama-3.1-8B) under different weight settings $(\alpha,\beta,\zeta)$.
The left column lists weights for correctness, detectability, and imperceptibility.
}
\label{tab:llama8b_b}
\end{table}

\begin{table*}[t]
\centering
\small
\setlength{\tabcolsep}{6pt}
\renewcommand{\arraystretch}{1.15}
\resizebox{\textwidth}{!}{%
\begin{tabular}{p{2.8cm}p{4.5cm}p{4.5cm}p{4.5cm}}
\toprule
\textbf{Category} & \textbf{Python} & \textbf{C++} & \textbf{Java} \\
\midrule
Keywords &
True, False, None, and, as, assert, async, await, break, class, continue, def, del, elif, else, except, finally, for, from, global, if, import, in, is, lambda, nonlocal, not, or, pass, raise, return, try, while, with, yield &
alignas, alignof, and, and\_eq, asm, auto, bitand, bitor, break, case, catch, class, compl, concept, const, consteval, constexpr, constinit, const\_cast, continue, co\_await, co\_return, co\_yield, decltype, default, delete, do, dynamic\_cast, else, enum, explicit, export, extern, false, for, friend, goto, if, inline, mutable, namespace, new, noexcept, not, not\_eq, nullptr, operator, or, or\_eq, private, protected, public, register, reinterpret\_cast, requires, return, sizeof, static, static\_assert, static\_cast, struct, switch, template, this, thread\_local, throw, true, try, typedef, typeid, typename, union, using, virtual, volatile, while, xor, xor\_eq, override &
abstract, assert, break, case, catch, class, const, continue, default, do, else, enum, extends, final, finally, for, goto, if, implements, import, instanceof, interface, native, new, null, package, private, protected, public, return, static, strictfp, super, switch, synchronized, this, throw, throws, transient, try, void, volatile, while, true, false \\
\midrule
Whitespace & space, \textbackslash n, \textbackslash t & space, \textbackslash n, \textbackslash t & space, \textbackslash n, \textbackslash t \\
\midrule
Types &
int, float, complex, str, bytes, bool, list, tuple, set, dict, NoneType &
int, float, double, bool, char, short, long, void, unsigned, signed, size\_t, ptrdiff\_t, wchar\_t, char8\_t, char16\_t, char32\_t &
byte, short, int, long, float, double, boolean, char, String, Object \\
\midrule
Delimiters &
(, ), [, ], \{, \}, `,', :, `.', ;, @, ->, ... &
(, ), [, ], \{, \}, `,', :, `.', ;, ->, ::, ... &
(, ), [, ], \{, \}, `,', :, `.', ;, @, ->, ::, ... \\
\midrule
Operators &
+, -, *, /, \%, **, //, =, ==, !=, >, <, >=, <=, +=, -=, *=, /=, \%=, //=, **=, \&, |, <<, >>, \textasciicircum, \textasciitilde &
+, -, *, /, \%, ++, --, =, ==, !=, >, <, >=, <=, \&\&, ||, !, \&, |, \textasciicircum, \textasciitilde, $<<$, $>>$, +=, -=, *=, /=, \%=, \&=, |=, \textasciicircum=, $<<=$, $>>=$, .*, ->* &
+, -, *, /, \%, ++, --, =, ==, !=, >, <, >=, <=, \&\&, ||, !, \&, |, \textasciicircum, \textasciitilde, $<<$, $>>$, $>>>$, +=, -=, *=, /=, \%=, \&=, |=, \textasciicircum=, $<<=$, $>>=$, $>>>=$, $\mathtt{>>>=}$ \\
\bottomrule
\end{tabular}%
}
\caption{Comparison of five categories of syntax elements across Python, C++, and Java.}
\label{tab:syntax_elements}
\end{table*}

\begin{table*}[t]
\centering
\small
\setlength{\tabcolsep}{3pt}
\renewcommand{\arraystretch}{1.12}
\begin{adjustbox}{max width=\textwidth}
\begin{tabular}{@{}lcccccccccccccccc@{}}
\noalign{\hrule height 0.8pt}
\multirow{2}{*}{Weight $(\alpha,\beta,\zeta)$} &
\multicolumn{4}{c}{MBPP+} &
\multicolumn{4}{c}{HumanEval+} &
\multicolumn{4}{c}{HEP-C++} &
\multicolumn{4}{c}{HEP-Java} \\
\cmidrule(lr){2-5}\cmidrule(lr){6-9}\cmidrule(lr){10-13}\cmidrule(lr){14-17}
& KGW & EWD & SWEET & \texttt{STONE}
& KGW & EWD & SWEET & \texttt{STONE}
& KGW & EWD & SWEET & \texttt{STONE}
& KGW & EWD & SWEET & \texttt{STONE} \\
\midrule
$(\tfrac{1}{3},\,\tfrac{1}{3},\,\tfrac{1}{3})$
& 0.775 & 0.819 & 0.787 & \textbf{0.848}
& 0.694 & 0.763 & 0.754 & \textbf{0.781}
& 0.730 & 0.750 & 0.735 & \textbf{0.780}
& 0.642 & 0.675 & 0.631 & \textbf{0.715} \\
$(0.5,\,0.25,\,0.25)$
& 0.706 & 0.739 & 0.716 & \textbf{0.778}
& 0.664 & 0.716 & 0.709 & \textbf{0.732}
& 0.692 & 0.706 & 0.697 & \textbf{0.741}
& 0.578 & 0.603 & 0.577 & \textbf{0.648} \\
$(0.25,\,0.5,\,0.25)$
& 0.789 & 0.856 & 0.807 & \textbf{0.881}
& 0.651 & 0.755 & 0.743 & \textbf{0.780}
& 0.703 & 0.733 & 0.711 & \textbf{0.768}
& 0.618 & 0.668 & 0.618 & \textbf{0.716} \\
$(0.25,\,0.25,\,0.5)$
& 0.830 & 0.863 & 0.838 & \textbf{0.883}
& 0.767 & 0.810 & 0.808 & \textbf{0.830}
& 0.796 & 0.811 & 0.796 & \textbf{0.833}
& 0.730 & 0.755 & 0.699 & \textbf{0.781} \\
\noalign{\hrule height 0.8pt}
\end{tabular}
\end{adjustbox}
\caption{\texttt{STEM} scores under different weight settings $(\alpha,\beta,\zeta)$.
The left column lists weights for correctness, detectability, and imperceptibility.
Bold indicates the best score in each row.}
\label{tab:STEM_weights}
\end{table*}

\section{Syntax Elements in Python, C++, and Java}
\label{app:example_of_syntax_elements}

Table~\ref{tab:syntax_elements} categorizes the syntactic elements 
into five groups: keywords, whitespace, types, delimiters, and operators. 
The selection of syntactic elements reflects the distinct characteristics 
of each programming language.

\begin{table*}[t]
\centering
\small
\setlength{\tabcolsep}{3pt}
\renewcommand{\arraystretch}{1.15}
\begin{tabular}{crcccccccc}
\noalign{\hrule height 0.8pt}
\multicolumn{2}{c}{\multirow{2}{*}{\shortstack{Entropy\\ Threshold}}} & 
\multicolumn{2}{c}{MBPP+} & 
\multicolumn{2}{c}{HumanEval+} &
\multicolumn{2}{c}{HEP-C++} & 
\multicolumn{2}{c}{HEP-Java} \\ 
\cmidrule(lr){3-4}\cmidrule(lr){5-6}\cmidrule(lr){7-8}\cmidrule(lr){9-10}
\multicolumn{2}{c}{} &
\multicolumn{1}{c}{\shortstack{Selected\\Tokens (\%)}} &
\multicolumn{1}{c}{\shortstack{Syntax\\Tokens (\%)}} &
\multicolumn{1}{c}{\shortstack{Selected\\Tokens (\%)}} &
\multicolumn{1}{c}{\shortstack{Syntax\\Tokens (\%)}} &
\multicolumn{1}{c}{\shortstack{Selected\\Tokens (\%)}} &
\multicolumn{1}{c}{\shortstack{Syntax\\Tokens (\%)}} &
\multicolumn{1}{c}{\shortstack{Selected\\Tokens (\%)}} &
\multicolumn{1}{c}{\shortstack{Syntax\\Tokens (\%)}} \\ 
\midrule
& 0.7 & 26.03 & 12.82 & 35.28 & 12.84 & 30.10 & 12.99 & 26.60 & 11.98 \\    
& 0.8 & 21.40 & 12.59 & 32.67 & 12.24 & 25.93 & 12.35 & 22.47 & 11.06 \\    
& 0.9 & \textbf{19.64} & \textbf{11.39} & \textbf{28.98} & \textbf{12.60} 
       & \textbf{22.65} & \textbf{11.79} & \textbf{18.52} & \textbf{11.68} \\    
& 1.0 & 16.59 & 11.39 & 24.84 & 11.83 & 18.96 & 10.56 & 16.13 & 11.15 \\    
& 1.1 & 14.94 & 10.30 & 23.22 & 11.81 & 15.71 & 9.82 & 14.39 & 10.91 \\    
\noalign{\hrule height 0.8pt}
\end{tabular}
\caption{
Token selection statistics of SWEET across four datasets.
\emph{Selected Tokens (\%)} denotes the proportion of tokens exceeding the entropy threshold out of all tokens,
and \emph{Syntax Tokens (\%)} represents the percentage of syntax-related tokens among those selected.
Bold values indicate the optimal entropy threshold used in SWEET.
}
\label{tab:sweet_watermark_token_selection}
\end{table*}

\section{Comprehensive Evaluation across \texttt{STEM} Weight Configurations}
\label{app:stem-grid}

Table~\ref{tab:STEM_weights} reports representative \texttt{STEM} variants 
using equal weights and three dimension-focused settings. 
We further explore all valid weight configurations where 
$\alpha,\beta,\zeta \in \{0.0,0.1,\dots,1.0\}$ and 
$\alpha+\beta+\zeta=1$, 
resulting in $66$ combinations. 
For each configuration, we compute the weighted \texttt{STEM} score and record 
the method with the highest value. 
Across all datasets, \texttt{STONE} performs consistently well throughout 
the entire weight space, ranking first in $97.0\%$ of configurations on MBPP+, 
$90.9\%$ on HumanEval+, $98.5\%$ on HumanEvalPack-C++, and $95.5\%$ on HumanEvalPack-Java. 
This consistency indicates that the advantage of \texttt{STONE} does not depend on 
specific weight choices but generalizes across diverse evaluation preferences, 
demonstrating robust performance under varying priorities of correctness, 
detectability, and imperceptibility.

\section{Syntax Token Coverage in SWEET}
\label{app:sweet_watermark_token_selection}

Table~\ref{tab:sweet_watermark_token_selection} presents an analysis of token selection 
by SWEET using two measures: 
(1) the proportion of selected tokens exceeding the entropy threshold among all generated tokens, 
and (2) the proportion of syntax-related tokens among those selected. 
We use the optimal settings~($\gamma=0.25$, $\delta=3.0$, entropy threshold $=0.9$)
and evaluate threshold values in the range $\{0.7,0.8,0.9,1.0,1.1\}$.
The results show an inverse relationship between the threshold and token coverage: 
lower thresholds select more tokens overall but include more syntax elements, 
whereas higher thresholds reduce syntax token inclusion. 
This pattern shows that SWEET effectively targets high-entropy regions 
but does not completely avoid syntax-related tokens.

Even under the optimal configuration, 
a notable portion of syntax tokens remains among those selected for watermarking. 
In the HumanEval+ dataset, syntax tokens account for 12.6\% of all selected tokens, 
including delimiters~(49.29\%), whitespace~(38.17\%), keywords~(9.44\%), types~(3.17\%), 
and operators~(2.78\%). 
This indicates that delimiters and whitespace dominate syntax-related selections, 
which directly affects code execution.

\begin{table}[h!]
\centering
\small
\setlength{\tabcolsep}{6pt}
\renewcommand{\arraystretch}{1.15}
\begin{tabular}{@{}lp{0.72\linewidth}@{}}
\toprule
\textbf{Category} & \textbf{Examples} \\
\midrule
Keywords    & def, or, import, for, True, assert, return, is, pass, None, False, in, not, if, from \\
Whitespace  & space, \textbackslash n \\
Types       & int, set, str \\
Delimiters  & . , : , \{, \}, [, ], (, ) \\
Operators   & *, \%, =, -, **, \&, \textasciicircum, /, +, | \\
\bottomrule
\end{tabular}
\caption{Examples of syntax tokens selected for watermarking in SWEET.}
\label{tab:sweet_watermark_token_selection_examples}
\end{table}

Table~\ref{tab:sweet_watermark_token_selection_examples} lists representative examples. 
Delimiters such as colons (`:') and commas (`,'), essential for code structure, 
can cause parsing errors when modified 
(e.g., omitting a colon in `for i in range(10):' leads to a failure). 
Although keywords and operators appear less frequently, 
their modification can still alter program behavior, 
for example, replacing `True' with `False' or `+' with `-'. 
Minimizing the inclusion of such syntax-critical tokens 
can make syntax-aware watermarking more reliable and less disruptive to functionality.

\section{Comparison with Non-watermarked Code}
\label{app:correctness-upperbound}

We evaluate the functional impact of watermarking by comparing the correctness of code generated with and without watermark insertion.
Non-watermarked code provides an empirical upper bound in our evaluation setting.
\texttt{STONE} achieves scores nearly identical to this bound, showing that watermark insertion has minimal influence on code execution.
Among all watermarking methods, \texttt{STONE} reduces correctness the least, which indicates that it embeds provenance information effectively while preserving the original program behavior.

\begin{table}[h!]
\centering
\small
\setlength{\tabcolsep}{6pt}
\renewcommand{\arraystretch}{1.1}
\begin{tabular}{lcc}
\toprule
Method & MBPP+ & HumanEval+ \\
\midrule
No watermark & 0.571 & 0.595 \\
KGW          & 0.499 & 0.573 \\
EWD          & 0.499 & 0.573 \\
SWEET        & 0.502 & 0.574 \\
\texttt{STONE} & \textbf{0.571} & \textbf{0.587} \\
\bottomrule
\end{tabular}
\caption{
Correctness comparison between non-watermarked and watermarked code.
Correctness is measured as pass@1 on MBPP+ and HumanEval+.
}
\label{tab:correctness_comparison}
\end{table}

\end{document}